\documentclass[aps,prl,amsfonts,amssymb,twocolumn,amsmath,preprintnumbers,nofootinbib,floatfix,
showpacs]{revtex4-1}
\usepackage[dvips]{graphics}
\usepackage{graphicx}
\usepackage{bm}
\begin{document}

\title{Long-range spin imbalance in mesoscopic superconductors under a Zeeman splitting}
\author{I. V. Bobkova}
\affiliation{Institute of Solid State Physics, Chernogolovka,
Moscow reg., 142432 Russia}
\affiliation{Moscow Institute of Physics and Technology, Dolgoprudny, 141700 Russia}
\author{A. M. Bobkov}
\affiliation{Institute of Solid State Physics, Chernogolovka,
Moscow reg., 142432 Russia}

\date{\today}

\begin{abstract}
We develop a theory of spin relaxation in Zeeman-splitted superconducting films at low temperatures. A new mechanism of spin relaxation, specific only for Zeeman-splitted superconductors is proposed. It can explain the extremely high spin relaxation lengths, experimentally observed in Zeeman-splitted superconductors, and their strong growth with the magnetic field. In the framework of this mechanism the observed spin signal is formed by the spin-independent nonequilibrium quasiparticle distribution weighted by the spin-split DOS. We demonstrate that the relaxation length of such a spin signal is determined by the energy relaxation length at energies of the order of the superconducting gap.
\end{abstract}
% insert suggested PACS numbers in braces on next line
\pacs{74.78.Na, 74.25.F-, 71.70.Ej}

\maketitle

Effective control of spin-polarized transport forms a basis of spintronic applications. In particular, it is very important to transmit spin signals over mesoscopic length scales. Usually at low temperatures the spin relaxation length is limited by elastic spin-flip at magnetic impurities or spin-orbit scattering processes. For example, it was shown in transport experiments \cite{jedema02,hubler12,quay13} that for Al thin films in the normal state the spin relaxation length $\lambda_N$ is of the order of $400-500$ nm. However, recently is has been reported for superconducting Al films that in the presence of a significant Zeeman splitting of the quasiparticle density of states (DOS) the spin signal can spread over distances of several $\mu$m \cite{hubler12,quay13,wolf13}. In these experiments the superconducting spin relaxation length exceeds considerably the superconducting coherence length, the normal-state spin relaxation length and the charge-imbalance length. It is also important that the relaxation length grows with the applied magnetic field. But a mechanism for such a long-distance spin relaxation is not understand yet.  

In the present paper we develop a theory of spin relaxation in Zeeman-splitted superconducting films at low temperatures. It is known that in the absence of the magnetic field (Zeeman splitting of the DOS) and at low temperatures the main mechanisms of the spin relaxation in superconductors are elastic spin flips by magnetic impurities and by spin-orbit interaction \cite{zhao95,morten04,morten05,poli08,yang10}. Here we show that it is unlikely that the experimentally observed long-distance spin relaxation is provided by such elastic spin-flip processes. Instead we suggest a new mechanism, which controls spin relaxation in Zeeman-splitted superconductors.

It is generally believed that the length of a spin signal spread is controlled by the characteristic length of any spin relaxation processes. We show that in the case of Zeeman-splitted superconductor this is not necessary so.  The spin relaxation length can be much larger than the length determined by fast elastic spin flip processes. The role of these elastic processes is only to rapidly relax the distribution function to the spin-independent value. The observed spin signal is formed by the spin-independent nonequilibrium quasiparticle distribution weighted by the spin-split DOS. We demonstrate that the relaxation length of such a spin signal is the energy relaxation length. This energy relaxation is provided by inelastic processes such as electron-electron and electron-phonon scattering. At low temperatures these inelastic processes are rather weak, so the corresponding spin relaxation lengths are large. Our theoretical estimates of the expected length scales for Al agree well with the experimental data \cite{hubler12,quay13,wolf13}. 

We show that the relaxation length for such a mechanism should grow with the magnetic field, as it is observed. The main  qualitative reason is the following. It is well-known that the scattering rates of inelastic processes are energy dependent at low temperatures: the electron-electron scattering rate $\tau_{e-e}^{-1} \sim \varepsilon^2$ and the electron-phonon scattering rate $\tau_{e-ph}^{-1} \sim \varepsilon^3$. The most essential energies for the considered here spin imbalance are of the order of the superconducting energy gap 
$\Delta$. The superconducting gap is suppressed by the magnetic field. This leads to the suppression of the characteristic energy scale important for the relaxation. Consequently, the characteristic scattering rate decreases and the relaxation length grows.

\begin{figure}[!tbh]
  %\centerline{\includegraphics[clip=true,width=2.5in]{Fig1_0.eps}}
           %\centerline{\includegraphics[clip=true,width=2.5in]{fig1b.eps}}
  \begin{minipage}[b]{0.5\linewidth}
     \centerline{\includegraphics[clip=true,width=1.7in]{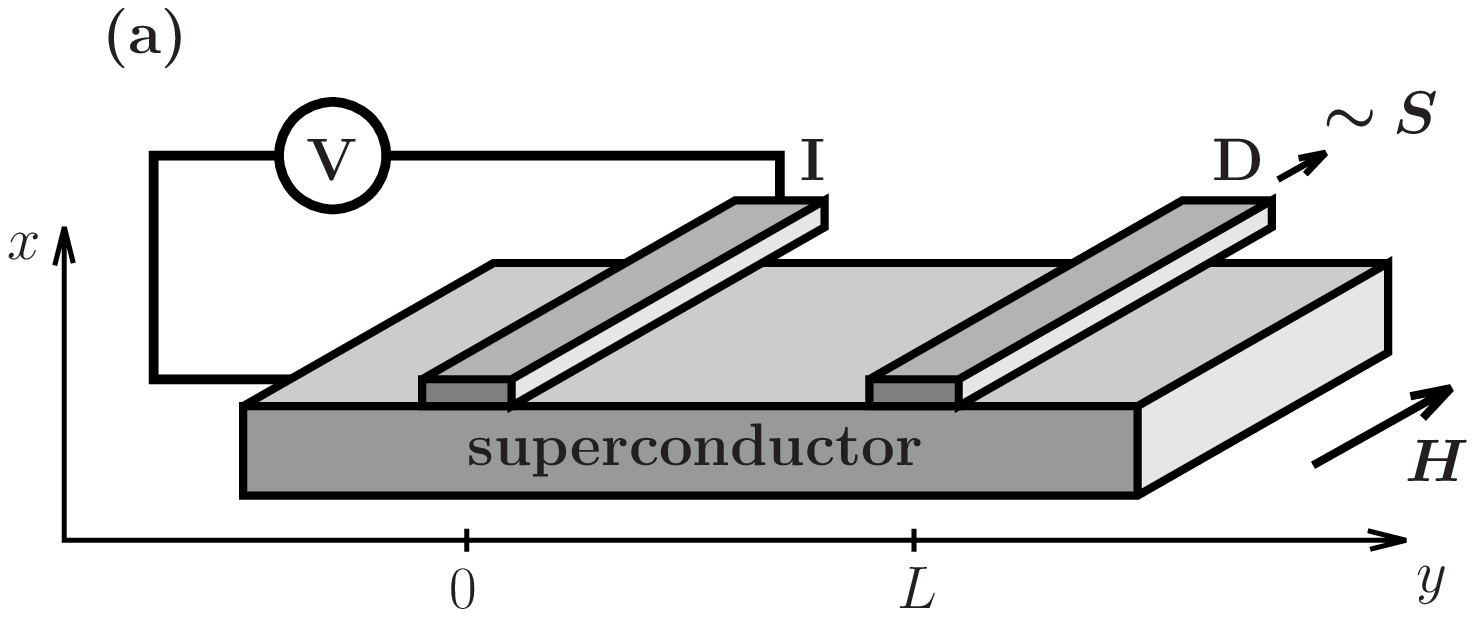}}
     \end{minipage}\hfill
    \begin{minipage}[b]{0.5\linewidth}
   \centerline{\includegraphics[clip=true,width=1.5in]{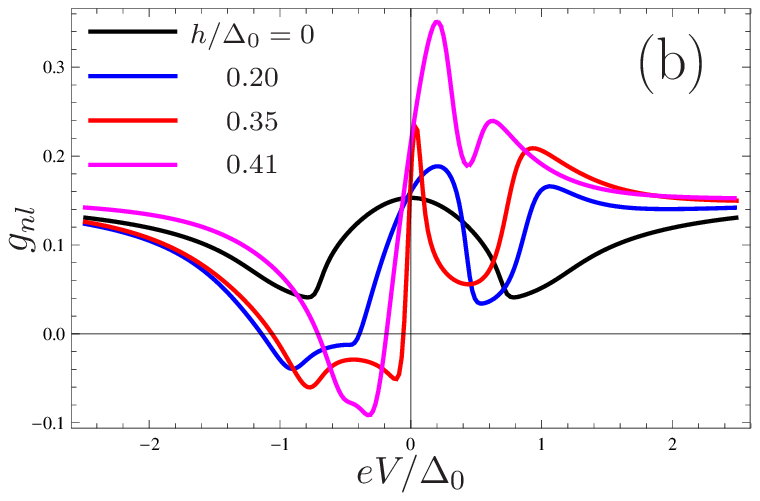}}
  \end{minipage}
\begin{minipage}[b]{0.5\linewidth}
     \centerline{\includegraphics[clip=true,width=1.5in]{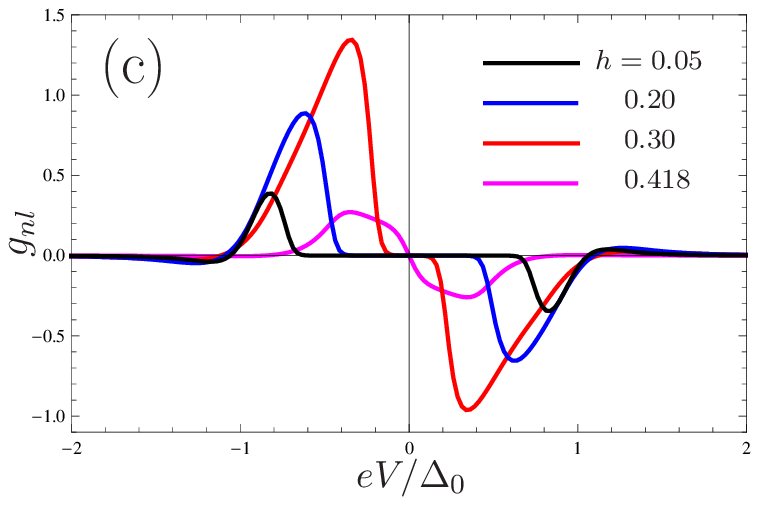}}
     \end{minipage}\hfill
   \begin{minipage}[b]{0.5\linewidth}
   \centerline{\includegraphics[clip=true,width=1.5in]{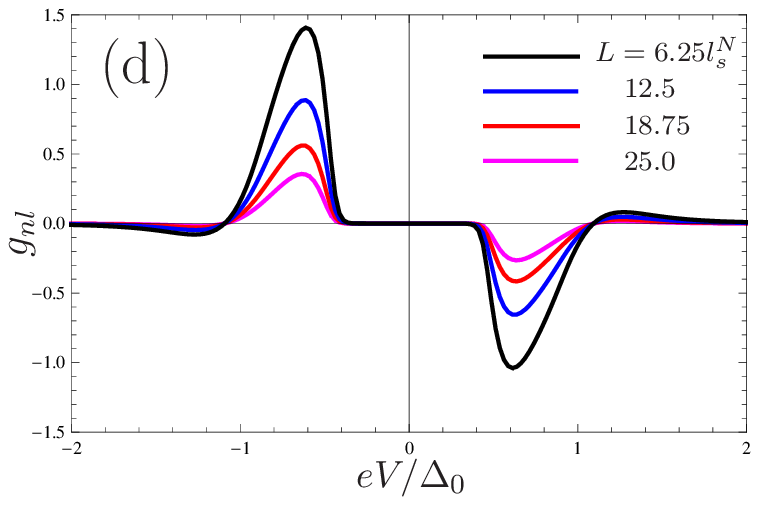}}
  \end{minipage}
  \begin{minipage}[b]{0.5\linewidth}
     \centerline{\includegraphics[clip=true,width=1.5in]{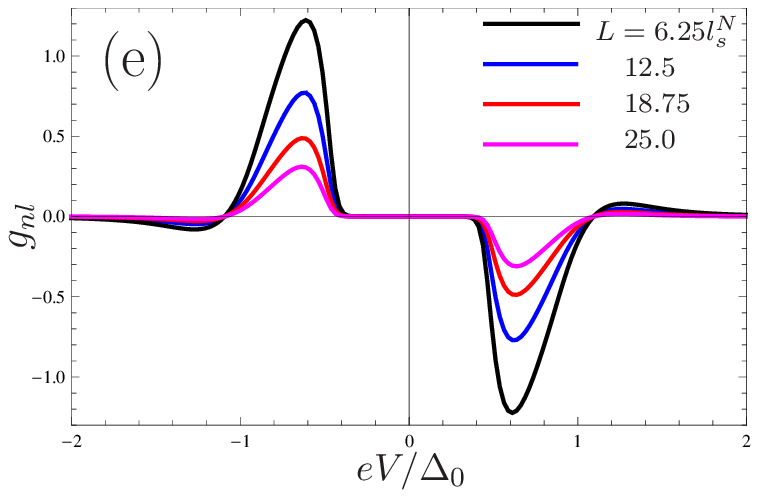}}
     \end{minipage}\hfill
    \begin{minipage}[b]{0.5\linewidth}
   \centerline{\includegraphics[clip=true,width=1.5in]{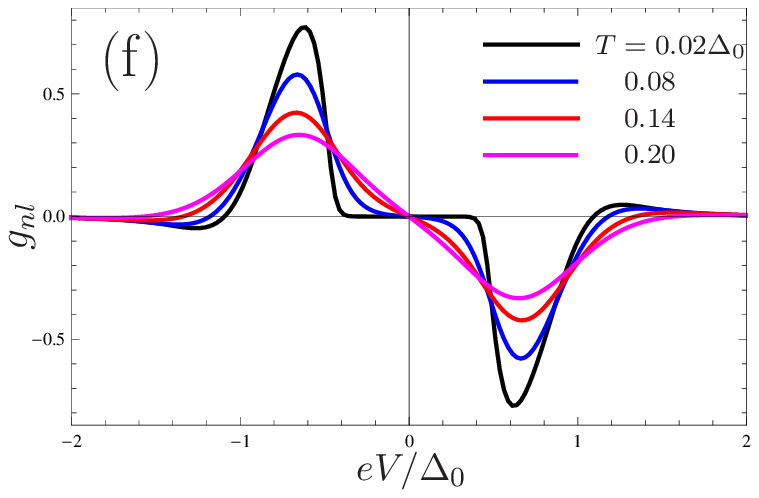}}
  \end{minipage}
   \caption{(Color online) (a) Sketch of the system under consideration. (b)-(c) nonlocal conductance as a function of $V$ for different magnetic fields. Panel (b) corresponds to elastic mechanisms of spin relaxation, $L=1.0l_s^N$. (c) $g_{nl}$ for the energy relaxation mechanism (see text), $L=12.5l_s^N$. (d) $g_{nl}$ as a function of $V$ for different $L$ for the energy relaxation mechanism. $h=0.20\Delta_0$. For panels (b)-(d) $P_I=0.2$. (e) $g_{nl}$ as a function of $V$ for the normal injector ($P_I=0$) and different $L$; $h=0.20\Delta_0$.  For panels (b)-(e) $T=0.02\Delta_0$. (f) $g_{nl}$ as a function of $V$ for the normal injector and different temperatures; $h=0.20\Delta_0$, $L=12.5l_s^N$. For all the panels $\tau_{so}^{-1}=\tau_{mi}^{-1}=0.015\Delta_0$.}   
\label{cond}
\end{figure}

Now we proceed to the calculation. Following the experiments \cite{hubler12,quay13,wolf13} we consider the system depicted in Fig.~\ref{cond}(a). It consists of a thin superconducting film (S) overlapped by the injector (I) and detector (D) electrodes. A current is injected into the superconducting film via I. This electrode can be normal or ferromagnetic. The detector electrode is ferromagnetic. The magnetic field is applied in plane of the film and is parallel to the ferromagnetic wires. It is interesting that the spin transport for misaligned magnetic field and injected spins was also studied recently \cite{silaev14}. In our study the quantization axis is chosen along the magnetic field. Both the injector and the detector are coupled to the film by tunnel contacts. 

In this case it can be shown that the non-local electric current, measured at the detector is proportional to $P_D S$. Here $P_D$ is the detector polarization and $S$ is the local nonequilibrium spin accumulation  in the film at the detector point. Further, we focus on this nonequilibrium spin accumulation $S$. This quantity can be written in terms of the Keldysh quasiclassical Green function as $S=\int \limits_{-\infty}^\infty d\varepsilon {\rm Tr} \left[\tau_3 \sigma_3 \left(\check g^K-\check g^K_{eq}\right)\right] /16$, where $\tau_i$ and $\sigma_i$ are Pauli matrices in the particle-hole and spin spaces, respectively. $\check g^K$ is the Keldysh component ($4 \times 4$ matrix) of the quasiclassical Green's function $\check g = \left( \begin{array}{cc} \check g^R & \check g^K \\ 0 & \check g^A \end{array} \right)$, where $\check g^{R(A)}$ are retarded and advanced Green's functions. $\check g^K_{eq}$ means the value of the Keldysh component in equilibrium. We assume the superconductor to be in the diffusive limit, so the matrix $\check g$ obeys the Usadel equation \cite{usadel,bergeret05}
\begin{eqnarray}
D \hat \partial_y (\check g \hat \partial_y \check g) + i \left[ \check \Lambda-\check \Sigma_{so}-\check \Sigma_{mi}-\check \Sigma_{in},\check g \right]=0
\label{usadel}
\enspace .
\end{eqnarray}
Here $\check \Lambda = \varepsilon \tau_3- h \sigma_3 \tau_3-\Delta i\tau_2$, $\varepsilon$ is the quasiparticle energy, $h=\mu_B H$ is the Zeeman field, $\Delta$ is the order parameter in the film and $D$ is the diffusion constant. $\hat \partial_y$ is a matrix in particle-hole space, accounting for the orbital suppression of superconductivity by the magnetic field. For a general matrix $\check G$ in particle-hole space $\hat \partial_y \check G =\partial_y \check G -(2ie/c)(Hx+A_0)\left[ P_{11}\check G P_{22} - P_{22} \check G P_{11} \right]$, where $P_{11(22)}=(1\pm\tau_3)/2$ and $x$ is the coordinate normal to the film. Eq.~(\ref{usadel}) should be supplemented by the normalization condition $\check g^2=1$.

The terms $\check \Sigma_{so}=\tau_{so}^{-1}(\bm \sigma \check g \bm \sigma)$ and $\check \Sigma_{mi}=\tau_{mi}^{-1}(\bm \sigma \tau_3 \check g \bm \sigma \tau_3)$ in Eq.~(\ref{usadel}) describe elastic spin relaxation processes of spin-orbit scattering and exchange interaction with magnetic impurities, respectively. The last term $\check \Sigma_{in}$ describes inelastic processes of energy relaxation.

We assume that the transparencies of the injector and detector interfaces are small, so that up to the leading (zero) order in transparency the retarded, advanced Green's functions and the order parameter take their bulk values in the presence of the magnetic field. The Green's functions can be represented in the form $\check g^R=g_0^R \tau_3 + g_t^R \sigma_3 \tau_3 + f_0^R i\tau_2 +f_t^R \sigma_3 i \tau_2$. It is convenient to use the following $\theta$-parametrization, which satisfies the normalization condition: $g_{0,t}^R=(\cosh \theta_+ \pm \cosh \theta_-)/2$ and $f_{0,t}^R=(\sinh \theta_+ 
\pm \sinh \theta_-)/2$. The advanced Green's functions can be found as $\check g^A=-\check g^{R*}$. Then one can obtain from Eq.~(\ref{usadel}) that $\theta_{\pm}$ obey the following equation:
\begin{eqnarray}
(\varepsilon \mp h)\sinh \theta_\pm +\Delta \cosh \theta_\pm +  \nonumber \\
Di \frac{e^2}{6c^2}H^2d^2 \cosh \theta_\pm \sinh \theta_\pm \pm 2i \tau_{so}^{-1}\sinh(\theta_+ - \theta_-)+\nonumber \\
2i\tau_{sf}^{-1}[\cosh \theta_\pm \sinh \theta_\pm +\sinh (\theta_+ +\theta_-)]=0
\label{theta}
\enspace .~~~~
\end {eqnarray}

Here the third term describes the orbital depairing of superconductivity. Usually this orbital deparing can be disregarded for thin films in parallel magnetic field. However, it can be estimated that for magnetic fields of the order of 1-2 T, which are applied in experiment, the orbital depairing can even exceed the other depairing factors (spin-orbit and magnetic impurity scattering). So, it cannot be neglected in Eq.~(\ref{theta}). In order to obtain Eq.~(\ref{theta}) we integrate the retarded part of Eq.~(\ref{usadel}) over the width $d$ of the film along the $x$-direction. $\Delta$ is calculated self-consistently. 

The term $\check \Sigma_{in}$, describing inelastic energy relaxation,  in principle, also enters Eq.~(\ref{theta}) as another depairing factor, but it is neglected because at low temperature it is small as compared to other depairing factors. It is important only for the calculation of the distribution function.

The normalization condition allows to write the Keldysh component as $\check g^K=\check g^R \check \varphi -\check \varphi \check g^A$, where $\check \varphi$ is the distribution function with the following general structure in particle-hole and spin spaces: $\check \varphi = (1/2)[\varphi_+^0 + \varphi_+^t\sigma_z + \varphi_-^0 \tau_z + \varphi_-^t \tau_z \sigma_z]$. Physically the distribution function $\varphi_-$ is responsible for the charge imbalance and $\varphi_+$ for the spin imbalance in the system. The components $\varphi_\pm^0$ describe the spin-independent part of the quasiparticle distribution, while $\varphi_\pm^t$ accounts for its spin polarization. In the equilibrium $\varphi_+^0=2 \tanh (\varepsilon/2T)$ and the other components of $\check \varphi$ are zero. Via the distribution function the nonequilibrium spin accumulation $S$ can be written as follows
\begin{equation}
S=\frac{1}{4} \int \limits_{-\infty}^\infty d \varepsilon \left( {\rm Re}[g_t^R](\varphi_+^0-2\tanh \frac{\varepsilon}{2T}) + {\rm Re}[g_0^R]\varphi_+^t \right)  
\label{spin}
\enspace .
\end{equation}    
It is worth to note here that for Zeeman-splitted superconductor the triplet part of the normal Green's function $g_t^R $ is nonzero, while it is vanishes for $h=0$. Due to this fact the nonequilibrium spin accumulation $S$ can be nonzero in the Zeeman-splitted superconductor even for the case of spin-independent quasiparticle distribution, that is for $\varphi_+^t=0$. In principle, for a Zeeman-splitted superconductor there is an equilibrium spin accumulation near the Fermi energy $S_{eq}=\frac{1}{2} \int \limits_{-\infty}^\infty d \varepsilon  {\rm Re}[g_t^R]\tanh (\varepsilon/2T) $. But we do not consider this quantity here because it does not contribute to the measured signal. It is interesting that the Zeeman splitting not only allows to get the spin signal on the basis of the spin-independent quasiparticle distribution, but it also provides a possibility to have highly spin-polarized electric current in normal metal/superconductor junctions \cite{giazotto08}. 

The equations for the distribution functions $\varphi_+^{0,t}$, entering Eq.~(\ref{spin}), can be derived from Eq.~(\ref{usadel}) and take the form
\begin{eqnarray}
D(\kappa_1 \partial_y^2 \varphi_+^0 + \kappa_2 \partial_y^2 \varphi_+^t)-\frac{\varphi_+^0-2 \tanh (\varepsilon/2T)}{\tau_\varepsilon}=0 \label{varphi1} \\
D(\kappa_2 \partial_y^2 \varphi_+^0 + \kappa_1 \partial_y^2 \varphi_+^t)-K\varphi_+^t-\frac{\varphi_+^t}{\tau_\varepsilon}=0
\label{varphi2}
\enspace .
\end{eqnarray} 
Here $\kappa_1=1+|g_0^R|^2+|g_t^R|^2-|f_0^R|^2-|f_t^R|^2$ and $\kappa_2=2{\rm Re}[g_0^R g_t^{R*}-f_0^R f_t^{R*}]$ account for the renormalization of the diffusion constant by superconductivity. $K=K_{so}+K_{mi}$ is responsible for the spin relaxation by elastic processes: spin-orbit scattering and spin-flip scattering by magnetic impurities, and
\begin{eqnarray}
K_{so(mi)}=8\tau_{so(mi)}^{-1}[ {\rm Re}({g_0^R}^2\mp {f_0^R}^2)+|g_0^R|^2\mp |f_0^R|^2- \nonumber \\
{\rm Re}({g_t^R}^2\mp {f_t^R}^2)-(|g_t^R|^2\mp |f_t^R|^2) ]
\label{K}
\enspace .~~~~~
\end{eqnarray}
The last terms in Eqs.~(\ref{varphi1})-(\ref{varphi2}) describe energy relaxation processes. We consider these processes in  relaxation time approximation. The renormalization of the energy relaxation time $\tau_\varepsilon$ by superconductivity is neglected. It has been reported in the literature \cite{kopnin09,moor09} that the main processes providing energy relaxation in Al at low temperatures are electron-electron scattering. So, we assume that $\tau_\varepsilon^{-1} \sim \gamma_{e-e}(T_c)\varepsilon^2/T_c^2$, where $\gamma_{e-e}(T_c) \sim 10^8 c^{-1}$ for Al.

We assume that the elastic spin-flip processes are much faster than the energy relaxation, that is $\tau_\varepsilon^{-1}/K \ll 1$. This assumption is in good agreement with the experimental situation \cite{hubler12,quay13}. Under this condition the solution of Eqs.~(\ref{varphi1})-(\ref{varphi2}) up to the leading order in the parameter $\tau_\varepsilon^{-1}/K $ takes the form
\begin{eqnarray}
\left(\begin{array}{c} 
\delta\varphi_+^0 \\
\varphi_+^t \end{array}\right)=\alpha \left(\begin{array}{c} 
-\frac{\kappa_2}{\kappa_1} \\
 1 \end{array}
\right)e^{-\lambda_s y}+
\beta \left(\begin{array}{c} 1 \\
\frac{\tau_\varepsilon^{-1}\kappa_2}{K\kappa_1} \end{array}\right)e^{-\lambda_\varepsilon y} 
\label{varphi_sol}
\enspace ,~~~~
\end{eqnarray}
where $\delta \varphi_+^0 =\varphi_+^0 -2\tanh \frac{\varepsilon}{2T} $, $\lambda_s^2=\kappa_1 K/D(\kappa_1^2-\kappa_2^2)$ and $\lambda_\varepsilon^2=\tau_\varepsilon^{-1}/D\kappa_1$. The first term in Eq.~(\ref{varphi_sol}) describes fast spin relaxation of the distribution function due to elastic spin-flip processes and the second corresponds to the slow energy relaxation of the approximately spin-independent part. 

Constants $\alpha$ and $\beta$ should be found from the appropriate boundary conditions for Eq.~(\ref{usadel}) at the injector/superconductor interface. These boundary conditions are to be obtained from the general Kupriyanov-Lukichev boundary conditions \cite{kuprianov88}, generalized for spin-filtering interfaces \cite{bergeret12}. In the considered case up to the leading order in the junction transparency we can neglect the superconducting proximity effect in the injector electrode. In this case the spectral function in it has a trivial spin and particle-hole structure: $\check g_I^{R,A}=\pm \tau_3$. Then the boundary conditions take the form   
\begin{equation}
\check g \hat \partial_y \check g = - \frac{\check G}{2\sigma_s}\left[ \check g, \check g_I \right]
\label{KL}
\enspace .
\end{equation}
If the injector is biased with respect to the superconductor by the voltage $V$, the Keldysh Green's function there takes the form $\check g_I^K=\tau_3 (\varphi_{I+}^0+\varphi_{I-}^0\tau_3)$, where
$\varphi_{I\pm}^0=\tanh[(\varepsilon-V)/2T]\pm\tanh[(\varepsilon+V)/2T]$. The tunnel interface between the injector and the superconductor is assumed to be spin-polarized with the conductance matrix $\check G=G_0+G_t\tau_3\sigma_3$. In the tunnel limit we consider the injected current polarization is mainly determined by the spin polarization of the tunnel conductance $P_I=G_t/G_0$. We take $P_I=0.2$ according to the experimental data \cite{hubler12}. $\sigma_s$ is the conductivity of the superconductor.

Boundary conditions for the distribution functions at $y=0$ are to be obtained making use of the Keldysh part of Eq.(\ref{KL}). Up to the leading (first) order in transparency of the I/S interface they take the form:
\begin{eqnarray}
\kappa_1 \partial_y \varphi_+^0+\kappa_2 \partial_y \varphi_+^t+2\frac{G_0}{\sigma_s}[{\rm Re}g_0^R](\varphi_{I+}^0-2\tanh \frac{\varepsilon}{2T}))+ \nonumber \\
2\frac{G_t}{\sigma_s}[{\rm Re}g_t^R]\varphi_{I-}^0=0 \enspace .~~~~~~~~  \label{varphi_bc1}\\
\kappa_1 \partial_y \varphi_+^t+\kappa_2 \partial_y \varphi_+^0+2\frac{G_0}{\sigma_s}[{\rm Re}g_t^R](\varphi_{I+}^0-2\tanh \frac{\varepsilon}{2T}))+ \nonumber \\
2\frac{G_t}{\sigma_s}[{\rm Re}g_0^R]\varphi_{I-}^0=0 
\label{varphi_bc2}
\enspace .~~~~~~~~~
\end{eqnarray} 
It is worth to note here that, while the distribution functions $\varphi_+$ and $\varphi_-$ obey the independent kinetic equations, they are coupled by the boundary conditions, if the interface barrier is spin-polarized, as it is seen from Eqs.~(\ref{varphi_bc1})-(\ref{varphi_bc2}). It is straightforward to find the constants $\alpha$ and $\beta$ from Eqs.~(\ref{varphi_bc1})-(\ref{varphi_bc2}). Up to the leading order in $\tau_\varepsilon^{-1}/K$:
\begin{eqnarray}
\beta=\frac{2}{\sigma_s \kappa_1 \lambda_\varepsilon}\left\{ G_0 [{\rm Re}g_0^R](\varphi_{I+}^0-2 \tanh \frac{\varepsilon}{2T})+ \right.~~~~~~~~ \nonumber \\
\left. G_t [{\rm Re}g_t^R]\varphi_{I-}^0\right\} ~~~~~~~~~~~~\label{beta} \\
\alpha=\frac{2\kappa_1}{\sigma_s (\kappa_1^2-\kappa_2^2) \lambda_s}\left\{ G_0 \left([{\rm Re}g_t^R]-\frac{\kappa_2}{\kappa_1}[{\rm Re}g_0^R]\right)\times \right. ~~~~~~~~~\nonumber \\
\left.(\varphi_{I+}^0-2 \tanh \frac{\varepsilon}{2T})+G_t \left([{\rm Re}g_0^R]-\frac{\kappa_2}{\kappa_1}[{\rm Re}g_t^R]\right)\varphi_{I-}^0\right\}
\label{alpha}
~~~~~~
\end{eqnarray}

Further, having at hand the spectral functions $g_{0,t}^R$, obtained from Eq.~(\ref{theta}) and the distribution functions, determined by Eq.~(\ref{varphi_sol}), we can calculate the nonequilibrium spin accumulation $S$ from Eq.~(\ref{spin}).

Now we turn to the discussion of the spin accumulation. At first we forget about  slow energy relaxation and study the elastic relaxation processes. We choose $\tau_{so}^{-1}+\tau_{mi}^{-1}=0.03\Delta_0$, where $\Delta_0 \equiv \Delta(h=0)$. Then the normal state spin relaxation length $l_s^N={\lambda^{N}_s}^{-1}=\sqrt{D/8(\tau_{so}^{-1}+\tau_{mi}^{-1})}$ corresponds to the experimental data for thin Al films \cite{jedema02,hubler12,quay13}. For the particular results presented here we also assume equal strengths of the spin-orbit and magnetic scatterings $\tau_{so}^{-1}=\tau_{mi}^{-1}$, but it does not influence qualitatively the results.

In Fig.~\ref{cond}(b) we demonstrate the nonlocal conductance calculated as $g_{nl}=dS/dV$. Different curves correspond to different applied magnetic fields. It is seen that the nonlocal conductance curves do not resemble the experimental results \cite{hubler12,quay13}. This takes place at any distance $L$ from the injector. The particular results, presented in Fig.~\ref{cond}(b), are calculated for 
$L=1.0l_s^N$. The main qualitative difference from the experimental results is that for the ferromagnetic injector the elastic relaxation gives as a symmetric in $V$, so as an anti-symmetric in $V$ components of the nonlocal conductance. In the experimental data the symmetric part is very small. For our calculated curves it is the dominating term. As it is seen from Eq.~(\ref{alpha}), it is always the case for a ferromagnetic injector, at least for the considered case of the tunnel junctions. In this case the nonlocal conductance can be fully antisymmetric in $V$ only for a normal injector.

\begin{figure}[!tbh]
  %\centerline{\includegraphics[clip=true,width=2.5in]{Fig1_0.eps}}
           %\centerline{\includegraphics[clip=true,width=2.5in]{fig1b.eps}}
  \begin{minipage}[b]{0.5\linewidth}
     \centerline{\includegraphics[clip=true,width=1.7in]{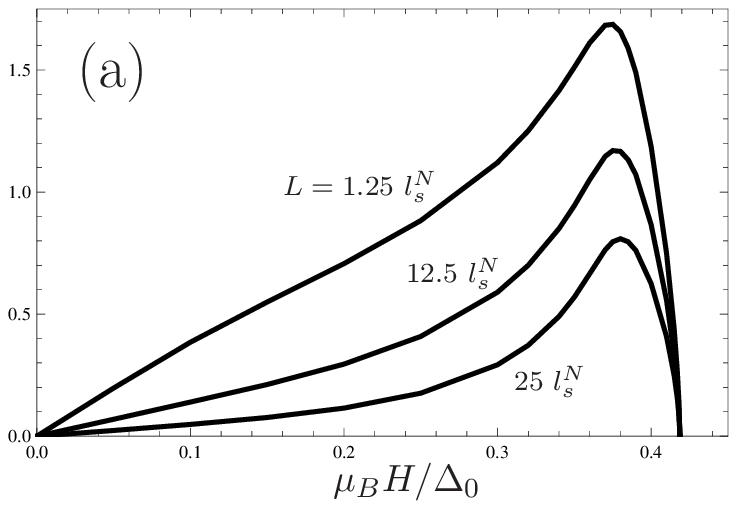}}
     \end{minipage}\hfill
    \begin{minipage}[b]{0.5\linewidth}
   \centerline{\includegraphics[clip=true,width=1.7in]{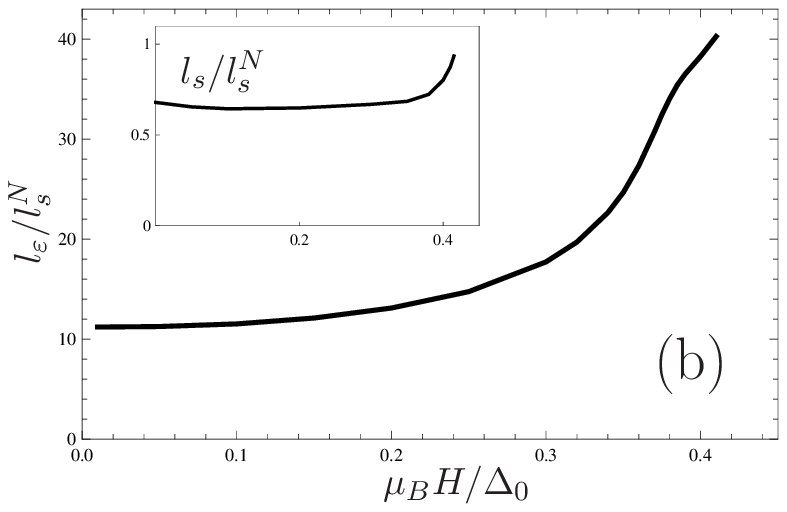}}
  \end{minipage}
%\begin{minipage}[b]{0.5\linewidth}
 %    \centerline{\includegraphics[clip=true,width=1.5in]{fig1c.eps}}
  %   \end{minipage}\hfill
   % \begin{minipage}[b]{0.5\linewidth}
   %\centerline{\includegraphics[clip=true,width=1.5in]{fig1d.eps}}
  %\end{minipage}
   \caption{(a) Peak area $S(V=-2\Delta_0)$ as a function of $H$ for different $L$. (b) Spin relaxation length $l_\varepsilon$ as a function of $H$. The inset represents the dependence of $l_s$ on $H$. $P_I=0.2$, $T=0.02\Delta_0$.}   
\label{length}
\end{figure}

The nonlocal spin signal $S$  decays exponentially as a function of distance $L$ from the injector point. In the insert to Fig.~\ref{length}(b) its decay length $l_s$ is plotted as a function of the magnetic field. Generally speaking, the decay length depends on the particular value of the voltage $V$ (quantitatively, not qualitatively). Our results are presented for $V=2.5\Delta_0$. We take $V$ to be considerably larger than $\Delta$ in accordance with the method used in \cite{hubler12} to estimate the value of the relaxation length. We can see from the insert to Fig.~\ref{length}(b) that $l_s$ is a nonmonotonous function of the field. After a slight initial decrease it grows at the large enough field, but this growth is too weak to account for the experimental data. More important thing is that the reasonable values of elastic scattering strength cannot be responsible for the large relaxation lengths $\sim $ several $\mu m$, observed in \cite{hubler12,quay13,wolf13}. The renormalization due to superconductivity only reduces the relaxation length with respect to its normal state value and $l_s \to l_s^N$ at large magnetic fields, when superconductivity is practically suppressed. 

So, on the basis of our analysis we can make the following conclusion. It is unlikely that the elastic relaxation processes is the mechanism of the slow spin relaxation, observed in \cite{hubler12,quay13,wolf13}. The main arguments are: (i) the shape of the nonlocal conductance does not resemble the experimental one for the ferromagnetic injector; (ii) the calculated values of $l_s$ are much smaller than the experimentally observed; (iii) the calculated growth with the applied field is too weak and $l_s$ is limited by $l_s^N$. Further we propose another mechanism of spin relaxation, which removes the most part of these disagreements. This can work only for a Zeeman-splitted superconductor, where the spin accumulation can be due to the ordinary spin-independent nonequilibrium quasiparticle distribution weighted by the spin-split DOS. 

Now let us include the weak energy relaxation processes in our study. From now we consider $L \gg l_s$, where the fast-decaying spin-dependent part of the distribution function is negligible. Then the distribution function is  nonequilibrium, but spin-independent. According to Eq.~(\ref{spin}), such a spin-independent distrubution can give the nonzero $S$ due to the Zeeman-splitted superconducting DOS. Figs.~\ref{cond}(c)-(f) represent the nonlocal conductance calculated for $L=12.5l_s^N$, where the distribution function is already practically spin-independent. Please note that we cannot say that "the elastic spin relaxation is not important here". It is these elastic processes that provide the fast relaxation of the distribution function to the spin-independent form. In the absence of such fast relaxation processes the distribution function would be spin-dependent in the Zeeman splitted superconductor due to energy relaxation, even for the case of normal injector.

The results, demonstrated in Figs.~\ref{cond}(c)-(f), are in good agreement with the experimental results. Figs.~\ref{cond}(c)-(d) represent the results for the ferromagnetic injector. Panel (c) demonstrates curves for different $H$ at fixed $L$ and panel (d) corresponds to different $L$ at fixed $H$. The plots are mainly anti-symmetric
in $V$, but the positive peak is slightly higher than the negative one. The peaks are fully anti-symmetric for a normal injector [see Fig.~\ref{cond}(e)-(f)]. The same feature is also observed experimentally \cite{hubler12,wolf13}. The temperature evolution of the conductance curves for the normal injector is plotted in Fig.~\ref{cond}(f) and is also in good agreement with the experimental findings. 

In Fig.~\ref{length}(a) the positive peak area (peak at $V<0$) is plotted as a function of $H$ for different $L$. The spin signal $S$ and, consequently, this peak area decay exponentially as a function of $L$. The decaying length $l_\varepsilon$ is plotted in Fig.~\ref{length}. 
It is seen that this inelastic spin relaxation length $l_\varepsilon$ shows a strong increase as a function of the magnetic field. It is worth to note that already at the smallest fields the value of $l_\varepsilon$ is considerably higher than the length of elastic relaxation. The physical reason for large increase of $l_\varepsilon$ with the field is twofold. 

(i) The energy-resolved spin relaxation length $l_\varepsilon(\varepsilon)$ is strongly suppressed for subgap energies. This suppression is reduced upon increase of the applied field, what results in some growth of $l_\varepsilon$ 
with $H$. However, this growth is not very essential, because the contribution of subgap energies to the overall signal is weighted by the subgap DOS. The subgap DOS is small, but nonzero due to the influence of the depairing factors, such as magnetic impurities, spin-orbit scattering and the orbital deparing. 

(ii) The second and, in fact, the main reason of $l_\varepsilon$ growth with the field is the suppression of the superconducting order parameter $\Delta$. The point is that the energies of the order of $\Delta$ make the most important contribution to the signal. The electron-electron scattering rate $\tau_\varepsilon^{-1} \propto \varepsilon^2$. Consequently, the effective scattering rate diminishes with the order parameter suppression. In the framework of this mechanism the upper limit for the spin relaxation length is the normal state energy relaxation length at a given temperature. If the energy relaxation is provided by the electron-phonon scattering $\tau_\varepsilon^{-1} \propto \varepsilon^3$, the qualitative picture is the same, but the more sharp increase of the relaxation length with the field is observed. 

For the considered model the main factor, which suppresses $\Delta$ upon rising the field is the orbital effect of the field. Although the film is thin, it cannot be disregarded here. We have estimated the corresponding deparing factor in Eq.~(\ref{theta}) as $D(e^2/6c^2)H^2d^2 =1.4 h^2/\Delta_0$.  This estimate agrees well with the value of the orbital deparing obtained in \cite{hubler10} by fitting the local conductance data. In such a case the critical field $H_c$ is mainly determined by the orbital deparing, and the contribution of the Zeeman deparing is rather small. The main role of the Zeeman term is to provide the splitted DOS.

Our calculated dependence $l_\varepsilon$ on the applied field manifests a strong growth when the field increases. This is in qualitative agreement with the experimental data. However, the particular shape of the calculated  $l_\varepsilon(H)$ does not agrees quantitatively with the measured ones \cite{hubler12, wolf13}.  One of possible reasons of such a discrepancy is an additional suppression of the superconducting order parameter in the vicinity of the ferromagnetic electrodes, which we do not take into account in our consideration. It results in inhomogeneities of the order parameter in the system. Then the space averaged "effective" order parameter should be suppressed by the field more smoothly than the homogeneous $\Delta$, assumed in our model. In its turn, it should provide more quantitative agreement with the experimental data. In particular, near the critical field $H_c$ the main energies, contributing to the relaxation, are determined by the width of the smeared coherence peak instead of $\Delta$. This effective width grows with the field, what can lead to a saturation or a decline of the relaxation length near $H_c$. However, a quantitative study of such an inhomogeneous problem is beyond the scope of this work. 

In conclusion, a theory of spin relaxation in Zeeman-splitted superconducting films at low temperatures is developed. It is suggested that the main mechanism, which determines the spin relaxation length in the Zeeman-splitted superconductors is the spin-independent energy relaxation. The role of faster elastic processes is only to relax the distribution function to the spin-independent form. In this framework the extremely high spin relaxation lengths, experimentally observed in Zeeman-splitted superconductors, and their strong growth with the magnetic field have natural explanations. In principle, the relaxation length can grow up to the normal state energy relaxation length for a given temperature. 

 %\begin{figure}[!tbh]
  %\centerline{\includegraphics[clip=true,width=3.5in]{fig4.eps}}
           %\centerline{\includegraphics[clip=true,width=2.5in]{fig1b.eps}}
  %\begin{minipage}[b]{0.5\linewidth}
   %  \centerline{\includegraphics[clip=true,width=1.5in]{fig1a.eps}}
    % \end{minipage}\hfill
    %\begin{minipage}[b]{0.5\linewidth}
   %\centerline{\includegraphics[clip=true,width=1.5in]{fig1b.eps}}
  %\end{minipage}
%\begin{minipage}[b]{0.5\linewidth}
 %    \centerline{\includegraphics[clip=true,width=1.5in]{fig1c.eps}}
  %   \end{minipage}\hfill
   % \begin{minipage}[b]{0.5\linewidth}
   %\centerline{\includegraphics[clip=true,width=1.5in]{fig1d.eps}}
  %\end{minipage}
  % \caption{}   
%\label{fig}
%\end{figure}

{\it Achnowledgments.} The authors are grateful to V.V. Ryazanov for useful discussions. The work was supported by Grant of the Russian Scientific Foundation No. 14-12-01290.

{\it Note added.} - We would like to note that there are two more preprints \cite{silaev14a, krishtop14},
which appeared practically simultaneously with ours and address the same effect. All the works are complementary
and devoted to different aspects of the long-range spin imbalance problem.

%\begin{figure}[!tbh]
 % \centerline{\includegraphics[clip=true,width=3.5in]{fig4.eps}}
            %\centerline{\includegraphics[clip=true,width=2.5in]{fig1b.eps}}
   %\begin{minipage}[b]{\linewidth}
    % \centerline{\includegraphics[clip=true,width=2.7in]{fig2a.eps}}
     %\end{minipage}\hfill
    %\begin{minipage}[b]{\linewidth}
   %\centerline{\includegraphics[clip=true,width=2.64in]{fig2b.eps}}
  %\end{minipage}
  % \caption{}   
%\label{distrib_dependence}
%\end{figure}

% Create the reference section using BibTeX:
%\bibliography{/users/tkm/howell/latexx/bibtexx/refs}

\end{document}